\documentclass[aip,amsmath,amssymb,reprint]{revtex4-2}

\usepackage{graphicx}% Include figure files
\usepackage{dcolumn}% Align table columns on decimal point
\usepackage{bm}% bold math
\usepackage{comment}
\usepackage[utf8]{inputenc}
\usepackage[T1]{fontenc}
\usepackage{mathptmx}
\usepackage{etoolbox}
\usepackage{float}
\usepackage{color,soul}

\makeatletter
\def\@email#1#2{%
 \endgroup
 \patchcmd{\titleblock@produce}
  {\frontmatter@RRAPformat}
  {\frontmatter@RRAPformat{\produce@RRAP{*#1\href{mailto:#2}{#2}}}\frontmatter@RRAPformat}
  {}{}
}%
\makeatother
\begin{document}

\preprint{AIP/123-QED}

\title[]{Anomalous shot noise in a bad metal $\beta$-tantalum}
% Force line breaks with \\
\author{Mateusz Szurek}
\affiliation{Department of Physics, Emory University, 400 Dowman dr., Atlanta, GA 30322.}%Lines break automatically or can be forced with \\
\email{mateusz.szurek@emory.edu}
\author{Hanqiao Cheng}%
\affiliation{Department of Physics, Emory University, 400 Dowman dr., Atlanta, GA 30322.}%Lines break automatically or can be forced with \\
\author{Zilu Pang}
\affiliation{Department of Physics, Emory University, 400 Dowman dr., Atlanta, GA 30322.}%Lines break automatically or can be forced with \\
\author{Yiou Zhang}
\affiliation{Department of Physics, Emory University, 400 Dowman dr., Atlanta, GA 30322.}%Lines break automatically or can be forced with \\
\author{John Bacsa}
\affiliation{Department of Chemistry, Emory University, 400 Dowman dr., Atlanta, GA 30322.}%Lines break automatically or can be forced with \\
\author{Sergei Urazhdin}
\affiliation{Department of Physics, Emory University, 400 Dowman dr., Atlanta, GA 30322.}%Lines break automatically or can be forced with \\

\date{\today}% It is always \today, today,

\begin{abstract}
We investigate the electronic shot noise produced by nanowires of $\beta$-Ta, an archetypal ``bad" metal with resistivity near the Ioffe-Regel localization limit. The Fano factor characterizing the shot noise exhibits a strong dependence on temperature and is suppressed compared to the expectations for quasiparticle diffusion, but hopping transport is ruled out by the analysis of scaling with the nanowire length. These anomalous behaviors closely resemble those of strange metal nanowires, suggesting that $\beta$-Ta may host a correlated electron liquid. This material provides an accessible platform for exploring exotic electronic states of matter.
\end{abstract}

\maketitle

%\section{\label{sec:level1}Introduction}

Anomalous metals are materials whose electronic properties cannot be described by the quasiparticle framework of conventional Fermi liquid theory.\cite{RevModPhys.91.011002,Chowdhury2022} ``Strange'' metals, which include high-temperature superconductors, exhibit a non-saturating linear dependence of resistivity $\rho$ on temperature $T$.\cite{PhysRevLett.127.086601,Greene2023} This behavior is thought to be related to electron correlations that underpin superconductivity (SC), potentially originating from incoherent Cooper pairing described as Bose liquid state.\cite{PhysRevB.91.115111,Hegg2021}

Materials known as ``bad'' metals exhibit high resistivity $\rho\gtrsim1.5\,\mu\Omega\cdot$m which nearly satisfies the Ioffe-Regel criterion for localization and slightly increases as $T$ decreases.\cite{Chowdhury2022} In the absence of resistivity divergence, their properties cannot be explained by single-particle localization.\cite{shklovskii1984variable} Instead, bad metals may be failed Mott insulators, where the insulating state is prevented by frustration of correlations or multi-orbital effects.\cite{Zhang2022} SC exhibited by some bad metals, such as TiN, $\beta$-W and Mo$_x$Ge$_{1-x}$, likely involves such correlations and may thus be unconventional.\cite{PhysRevB.102.060501,Bastiaans2021,PhysRevB.105.L060503} Bad metals $\beta$-Ta and $\beta$-W are among the most efficient sources of spin current based on the spin Hall effect (SHE).\cite{Liu555} A correlated state of these materials may challenge the single-particle interpretation of this effect. These possibilities warrant studies of the mechanisms of bad metallicity.

\begin{figure}[H]
\includegraphics[width=\columnwidth]{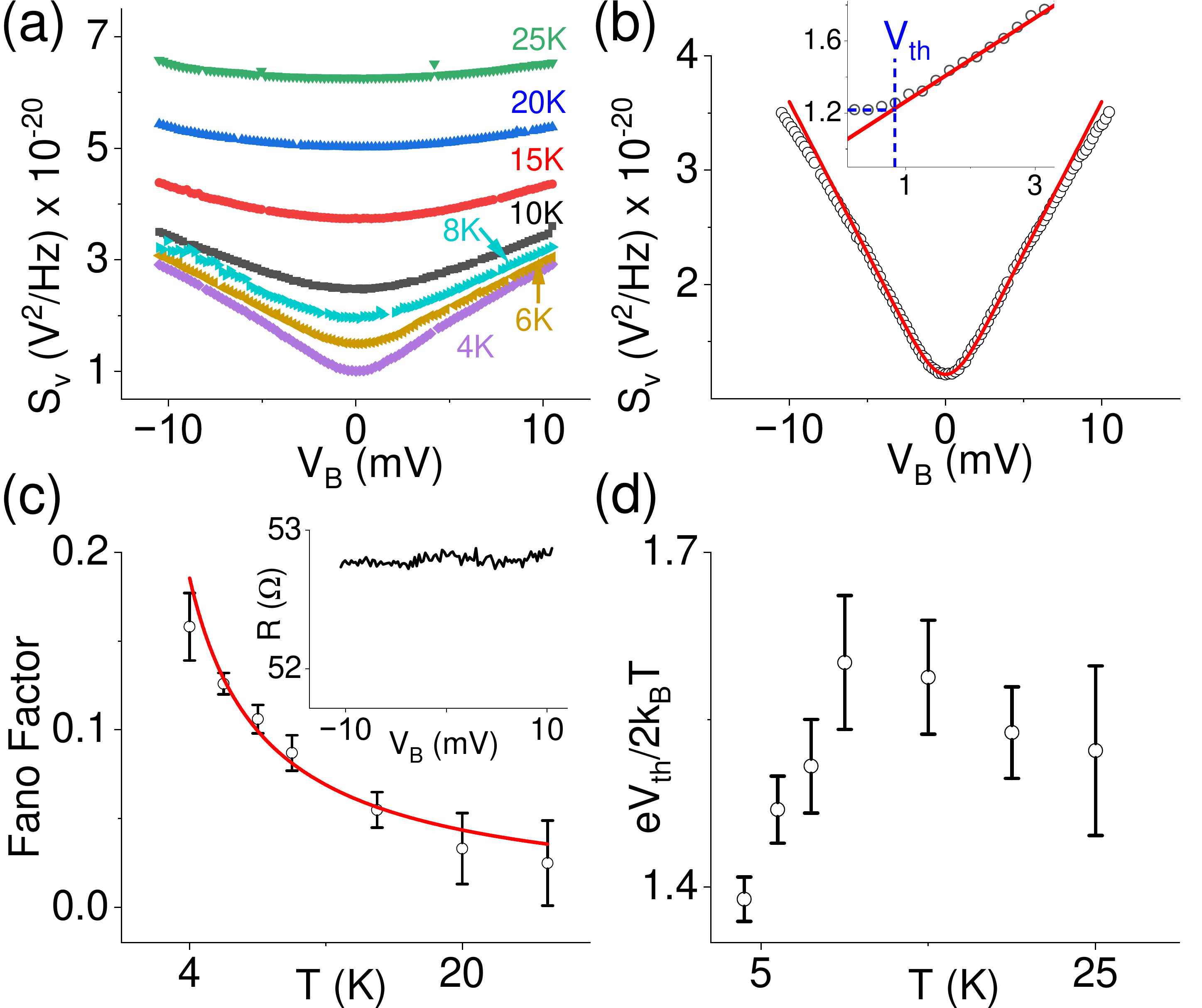}
\caption{\label{fig:2} Results for the $\beta$-Ta nanowire with $L=100$~nm. (a) Noise vs $V_B$, at the labeled $T$. (b) Noise vs $V_B$ at $4.2$~K, with fitting of data for $|V_B|<5$~mV using Eq.~(\ref{eq:SN_general}). Inset: $V_{th}$ as the intercept between SN and Johnson noise regimes. (c) $F$ vs $T$, curve is a fit with hyperbola. Inset: $R$ vs $V_B$ at $4.2$~K. (d) $eV_{th}/2k_BT$ vs $T$.}
\end{figure}

Measurements of electronic shot noise (SN) $-$ the noise produced by metallic or tunneling junctions due to the discrete nature of charge carriers $-$ have emerged as a powerful tool for studying correlations.\cite{SNCorrelations,ShotNoiseStrangeMetals,chen2023shot,PhysRevB.110.L081102,SNYiou} Voltage noise produced by metallic nanowires with a length $L$ smaller than the electron-phonon (e-ph) scattering length $l_{e-ph}$ can be described by

\begin{equation}\label{eq:SN_general}
 S_V = 2Fe{\frac{dV}{dI}}(V\coth \frac{V}{V_{th}}-V_{th})+4k_BT{\frac{dV}{dI}},
\end{equation}
where $e$ is the electron charge, $V_B$ is the bias voltage, $R$ is resistance, and the thermal broadening $V_{th}(T)$ determines the crossover from thermal to shot noise (SN) regimes [inset in Fig.~\ref{fig:2}(b)].\cite{ShotNoiseBible} At $V_B\ll V_{th}$ this equation describes Johnson noise $S_{V,J}=4k_BTR$, and at $V_B\gg V_{th}$ $-$ a linear dependence on bias expected for SN with the slope given by the Fano factor $F=S_{V}/2eV_BR|_{V_B\gg k_BT}$.

Both $F$ and $V_{th}$ depend on the mechanism of charge transport. Diffusive single-electron transport in a wire shorter than the electron-electron scattering length $l_{e-e}$ gives $F=1/3$ and the thermal broadening factor $B\equiv eV_{th}/2k_BT=1$.\cite{beenakker1992suppression,SNDirtyMetal} For $L>l_{e-e}$, electron thermalization results in $F=\sqrt{3}/4$ and $B=2/\sqrt{3}$.\cite{HotElectrons,SNelectronElectron,note_hot_e} At $L>l_{e-ph}$ the SN is suppressed by e-ph interaction, resulting in nonlinear dependence $S_V\propto V_B^{2/5}$.\cite{PhysRevB.49.5942}

Strange metal nanowires showed suppressed SN with $F<1/3$, which was attributed to a continuous charge fluid of correlated electrons.\cite{chen2023shot} SN suppression in a semimetal close to metal-insulator transition was explained by correlated electron hopping and described by $F=(L_0/L)^\beta$, where $L_0$ is the hopping length and $\beta=0.5-1$.\cite{kinkhabwala2006numerical,SNYiou} These two correlated states are differentiated by thermal broadening, $B=1$ in the former and a significantly larger $B=1/F$ in the latter case. 

\begin{figure}[H]
\includegraphics[width=\columnwidth]{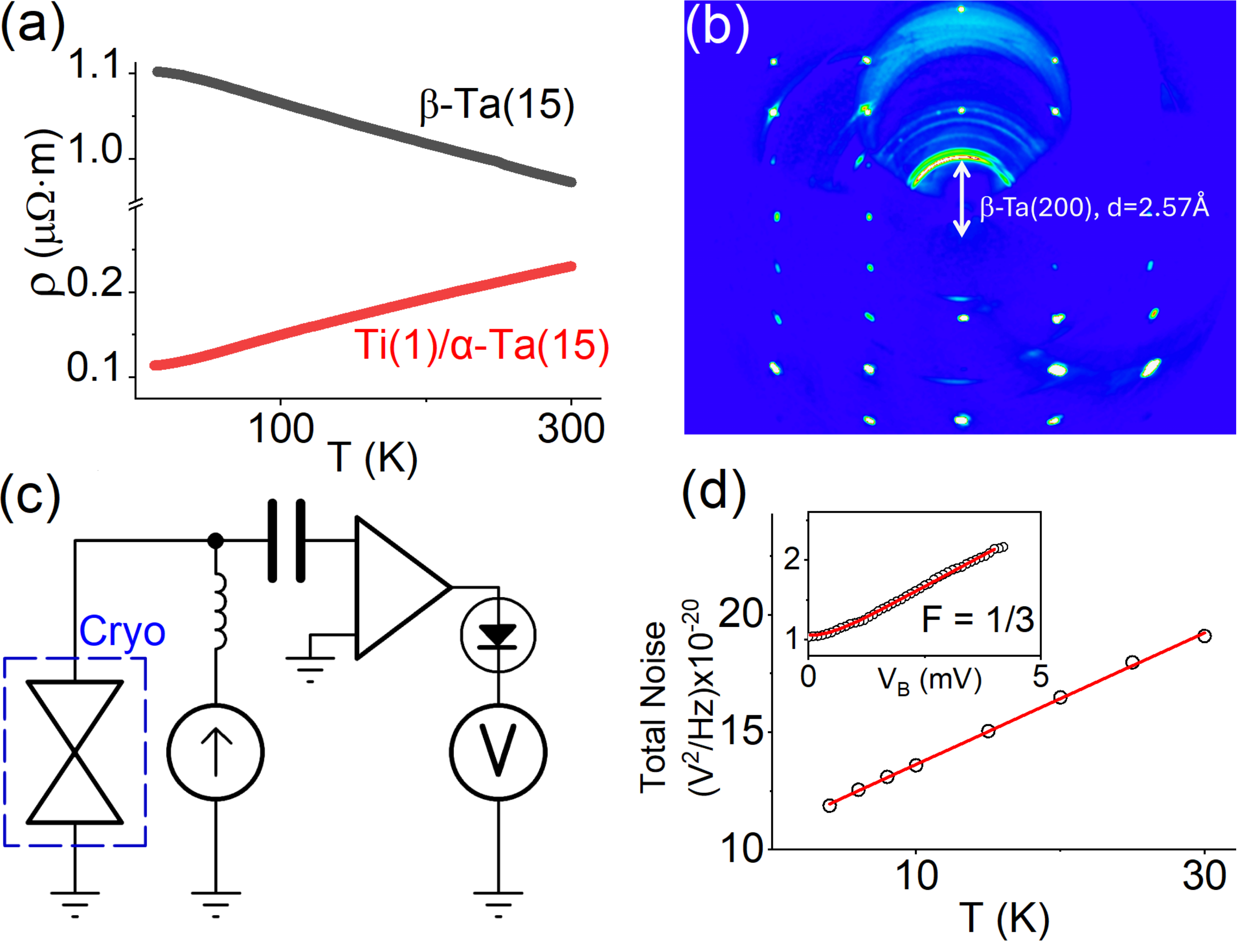}
\caption{\label{fig:1} (a) $\rho$ vs $T$ for $\beta$-Ta(15) and Ti(1)$\alpha$-Ta(15). (b) Laue diffraction pattern of $\beta$-Ta(35).  (c) Schematic of the noise measurement setup. (d) Dependence of Johnson noise on $T$ for a Au(10) nanowire with $L=500$~nm, before background noise subtraction. Inset: noise vs bias at $T=4.2$~K and fitting using Eq.~(\ref{eq:SN_general}) with $F=1/3$, $B=1$.}
\end{figure}

%\section{Results}

We report the observation of anomalous SN inconsistent with the Fermi liquid picture in nanowires of a bad metal $\beta$-Ta. The nanowires were fabricated from Ta films sputtered on sapphire (Al$_2$O$_3$) substrates in a vacuum chamber with a base pressure of $5\times10^{-9}$~Torr, in $4$~mTorr of ultrahigh-purity Ar. Nanometer-thick Ta films form predominantly the $\beta$ phase, while the $\alpha$ phase becomes dominant in thicker films.\cite{bTaProduction} We developed a deposition procedure to produce a single phase. $\alpha$-Ta films, used as a reference, were obtained using a Ti(1) buffer layer [the number in parentheses indicates thickness in nanometers], while the $\beta$ phase was achieved by exposing the substrate to an Ar ion beam with $300$~eV ion energy for a few seconds before Ta deposition. Since Al$_2$O$_3$ is very hard, this process removes a negligible amount of substrate material. The $\beta$ phase is likely stabilized by the loose surface bonds produced by Ar ions. Further studies are needed to establish the mechanism. Note that both $\alpha$- and $\beta$-Ta were deposited under identical conditions from the same source, ruling out impurity scattering as the origin of bad metallicity.

The procedure was verified by measurements of the resistivity of extended films, Fig.~\ref{fig:1}(a). For films grown on the Ti$(1)$ buffer, $\rho$ decreases with decreasing $T$, consistent with the Fermi liquid state of $\alpha$-Ta. In contrast, for films grown directly on Al$_2$O$_3$, $\rho$ is an order of magnitude larger and shows a small increase with decreasing $T$, as expected for a bad metal. 
%Similar results were obtained for films with thicknesses ranging from $5$ to $35$~nm, confirming the stability of this process. The anomalous temperature dependence was also confirmed for the nanowires.

The Laue diffraction pattern of the Ta film grown directly on Al$_2$O$_3$ is dominated by an intense $\beta$-Ta[200] ring, a diffuse $\beta$-Ta[400] ring and the Al$_2$O$_3$ peaks, Fig.~\ref{fig:1}(b). The identification of three additional low-intensity diffraction rings is challenging due to the complexity of $\beta$-Ta structure.\cite{bTaStructure,Magnuson2019} None of them correspond to the dominant bcc [110] reflection of $\alpha$-Ta.

The $\beta$-Ta(10) films used for the nanowires were capped with Au(5) to protect from oxidation. The films were patterned into nanowires by ion milling through a polymer mask, followed by the deposition of Ti(1)/Au(100) electrodes. Finally, the Au(5) was removed from the surface by Ar ion milling, and a capping AlO$_x$(2) layer was deposited to protect from oxidation. Ohmic contact was confirmed by negligible dependence of sample resistance on bias, inset in Fig.~\ref{fig:2}(c).

Noise measurements were performed using a microwave (mw) cryostat with a base temperature of $4.2$~K [see schematic Fig.~\ref{fig:1}(c)], which allows for quick data acquisition when compared to low-frequency noise measurement approaches. To maintain approximate impedance matching with the mw line, the width $w$ of the nanowires was scaled as $w=3L$, resulting in sample resistance $R\approx40-65\Omega$. The large width of the nanowires minimizes possible effects of edge roughness that can lead to reduced shot noise. The noise produced by the samples was separated from the dc bias by a bias tee and amplified by ultralow noise amplifiers ($3\times$ Qorvo QPL7547 in series) with a total gain of $10^8$ at a frequency $f_0=350$~MHz and a $3$~dB bandwidth of $50$~MHz. This narrow passband effectively filtered out flicker noise and electromagnetic interference (EMI), as was confirmed by the analysis of noise spectra. The signal was rectified by a mw diode detector (Omni Spectra 20760 Schottky Detector) and digitized by an A/D converter. Occasional outlier data points, attributed to spikes of broadband EMI, were removed. 

The setup was calibrated by the temperature dependence of Johnson noise, as illustrated in Fig.~\ref{fig:1}(d) for an $500$~nm-long  Au nanowire used as a calibration sample. The calibration accounted for the sample lead resistance $R_L=2.5\,\Omega$ and variations of $R$. The total resistance of the mw line was less than $0.5$~Ohm, resulting in negligible contribution to noise variations from its cryogenic section. We used a separate four-probe measurement to estimate that the total contact resistance between the Ta nanowires and the Au leads was no more than 0.2~$\Omega$, negligible compared to the nanowire resistance. The background produced by room-temperature circuit elements was subtracted. The calibration was verified by measurement of the bias dependence of SN produced by the Au nanowire, which yielded $F=1/3$ at $4.2$~K consistent with single-particle diffusion [inset in Fig.~\ref{fig:1}(d)]. 

The results for the $100$~nm $\beta$-Ta nanowire are summarized in Fig.~\ref{fig:2}. At $4.2$~K, the noise exhibits a linear bias dependence at $|V_B|<5$~meV, with thermal broadening at $|V_B|\lesssim1$~meV, inset in panel (b). These data are well approximated by Eq.~(\ref{eq:SN_general}), as shown by the fitting in Fig.~\ref{fig:2}(b). At $|V_B|\approx5$~mV, the noise falls below the linear dependence, consistent with the effects of  electron-phonon scattering at large bias, expected to result in a crossover to $S_V\propto V_B^{2/5}$.\cite{Henny1997,SNDiffusive}

The results of fitting with Eq.~(\ref{eq:SN_general}) of temperature-dependent data are summarized in Figs.~\ref{fig:2}(c),(d). The Fano factor scales approximately inversely with $T$, decreasing from $0.16\pm0.02$ at $T=4.2$~K to $0.03\pm0.03$ at $25$~K. Meanwhile, the thermal broadening factor remains almost constant, $B=1.4-1.6$. The relative uncertainties increase with $T$ due to thermal broadening and reduced dependence on bias. The dependence of $B$ on temperature is not expected from single-electron models, and may provide independent evidence for the correlation effects. However, because of the strong covariance between $F$ and $B$ in fitting with Eq.~(1), we cannot reliably conclude that the variations of the fitted value of $B$ are not caused simply by the limitations of the fitting and the variations of the Fano factor.
%At $25$~K, the uncertainty of $F$ becomes equal to its value. %Although the fitting uncertainty of $B$ remains modest, its value becomes unreliable due to the covariance of the two parameters.

\begin{figure}[H]
\includegraphics[width=\columnwidth]{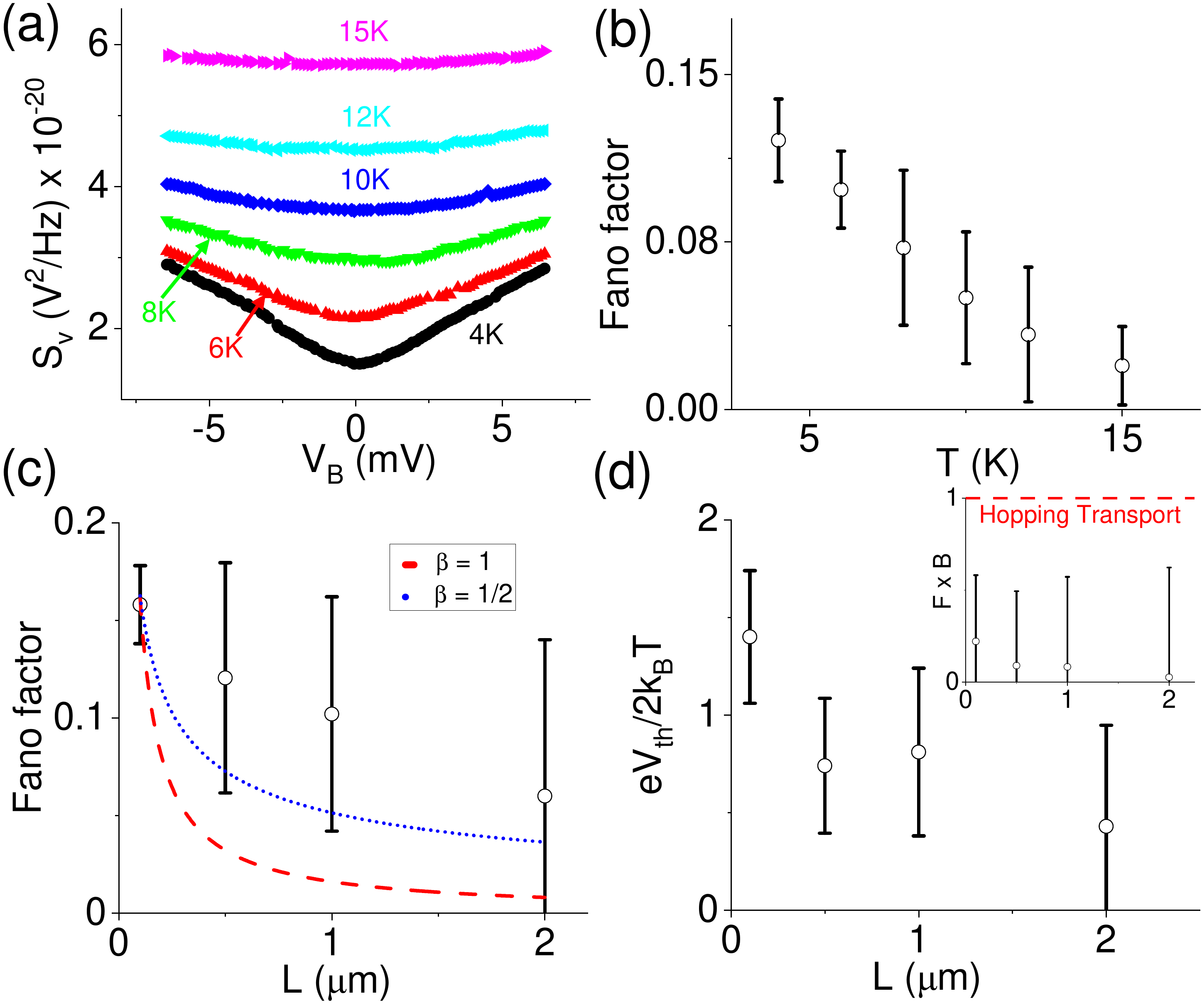}%
\caption{\label{fig:3} (a) Noise vs bias for $L=500$~nm, at the labeled $T$. (b) Fano factor vs $T$ for the sample in (a). (c),(d) Symbols: Fano $F$ (c) and thermal broadening (d) factors vs $L$, at $T=4.2$~K. Curves in (c): best fits with $F = (L_0/L)^\beta$ expected for hopping transport, with $\beta=1$ (dashed curve) and $\beta=0.5$ (dotted curve). Inset in (d): $F*B$ vs $L$ with a reference line $F*B = 1$ expected for hopping transport.} 
\end{figure}

Similar evolution of noise characteristics with $T$ was observed for longer nanowires, as shown in Figs.{~\ref{fig:3}}(a),(b) for $L=500$~nm. The dependence of noise on bias exhibited a well-defined linear regime at modest $V_B$, enabling fitting with Eq.~({\ref{eq:SN_general}}) to determine $F$ and $B$. We now focus on the results for the base temperature $T=4.2$~K, where small thermal broadening gives the lowest uncertainty of parameters extracted from the fitting. The Fano factor decreases as $L$ is increased from $100$~nm to $2\,\mu$m by a modest amount comparable to its uncertainty [Fig.{~\ref{fig:3}}(c)]. These results place significant constraints on the possible mechanisms of charge transport and the electronic state of $\beta$-Ta, as discussed below.

At large bias, e-ph scattering is expected to result in downcurving of the bias dependence due to e-ph scattering.\cite{PhysRevB.49.5942} This effect allows us to evaluate the possibile role of e-ph scattering in our measurements. In the hot electron approximation, which may be justified by the strong electron-electron interaction expected to thermalize the single-particle distribution, the electron temperature $T_e$ is given by the diffusion equation~\cite{PhysRevB.49.5942,Henny1997}
\begin{equation}\label{eq:diffusion}
\frac{\pi^2}{6}\frac{\partial^2 T_e^2}{\partial x^2}=-(\frac{eV_B}{k_BL})^2+\Gamma (T^5_e-T^5_{ph}),    
\end{equation}

where $\Gamma$ is the e-ph coupling constant, and $T_{ph}$ is phonon temperature. Equation~({\ref{eq:diffusion})} is generally applicable as long as the hot electron approximation ($L \gg L_{e-e}$) is justified. The electron-phonon scattering length can also be estimated by equating the two terms on the right-hand side of Eq.(2), which gives $L_{e-ph} \sim eV/k_BT^{2.5}\Gamma^{0.5}$. For small $\Gamma$ and sufficiently small bias, this equation describes hot electron SN regime with $F=\sqrt{3}/4$. At large $\Gamma$, the dominant e-ph scattering results in $S_V\propto V^{2/5}$ which can appear as suppressed SN. As temperature increases, $\Gamma$ generally increases, which may account for the further suppression of shot noise at higher temperature.

\begin{figure}[H]
\includegraphics[width=\columnwidth]{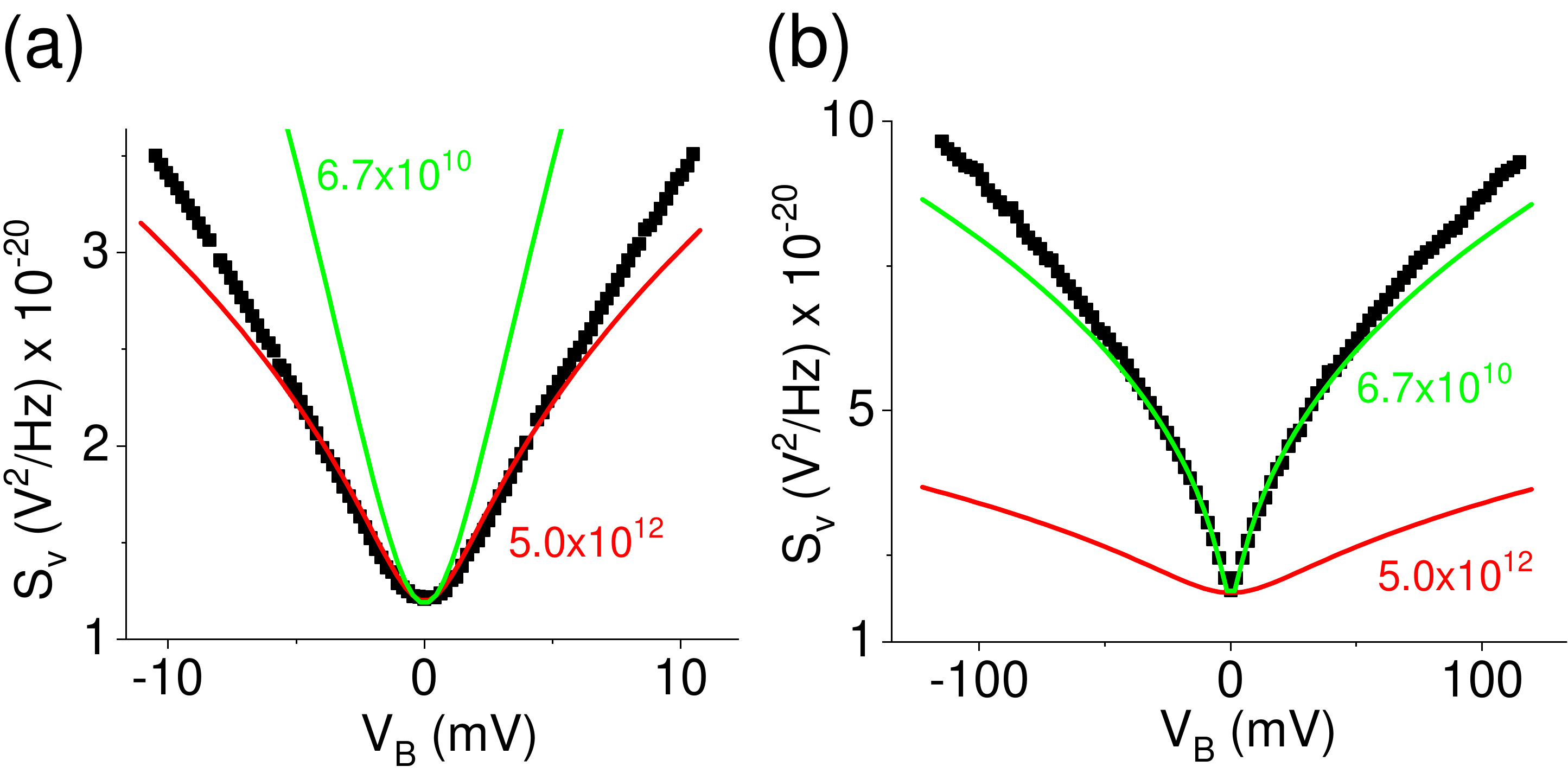}%
\caption{\label{fig:4} Symbols: Noise vs bias for $L=100$~nm (a) and $L=1$~$\mu$m (b) at $T=4.2$~K. Curves: calculations based on Eq.~(\ref{eq:diffusion}) using the labeled values of $\Gamma$ in units of $K^{-3} m^{-2}$.}
\end{figure}

We solved Eq.~(\ref{eq:diffusion}) numerically, with the values of $\Gamma$ adjusted to fit the low-bias slope of the observed noise, as shown in Fig.~\ref{fig:4}. For $L=100$~nm, the optimal value is $\Gamma=5.0\times10^{12}$~K$^{-3}$m$^{-2}$. For $L=1$~$\mu$m, the corresponding value is $\Gamma=6.7\times10^{10}$~K$^{-3}$m$^{-2}$, two orders of magnitude smaller. The value obtained for the $100$~nm wire cannot account for the dependence observed for the $1\,\mu m$ wire, and vice versa, as shown in Fig.~\ref{fig:4}. Even for the values of $\Gamma$ well-approximating the low-bias dependence, calculations overestimate the downcurving of the data. These large discrepancies suggest that e-ph scattering is not the mechanism of SN suppression.

%\section{Discussion}
We now discuss the possible mechanisms of the observed noise characteristics. Since $\beta$-Ta is close to the Ioffe-Regel limit, one may assume that transport is mediated by hopping of almost localized electrons, which is expected to result in suppressed $F$.\cite{SNYiou} For single-electron hopping, F=$(L_0/L)^\beta$ and $B=1/F$, where $L_0$ is the hopping length and $\beta=0.5-1$. While $\beta$~=~1 does not give a good fit for the length dependence, $\beta$~=~0.5 gives a marginally acceptable fit due to the large errors in the Fano factor, curves in Fig.{~\ref{fig:3}(c).} However, thermal broadening characterized by the product $F*B$ is much smaller than $1$ expected for hopping (inset in {Fig.\ref{fig:3}}(d)). Therefore, shot noise suppression due to hopping transport can be ruled out, which confirms that disorder-driven single-electron localization is not relevant to $\beta$-Ta.

The possibility that SN is suppressed by e-ph interaction is contradicted by the large discrepancies of the observed dependence of noise on length with calculations, as discussed above. We provide an additional qualitative argument against this mechanism. If local dissipation into phonons and then into the substrate is dominant, one can neglect the left-hand side of Eq.~(\ref{eq:diffusion}). In this limit, the electron temperature is determined by the electric field $E=V_B/L$ instead of the total voltage bias. For a given $V_B$, electric field $E$ decreases $20$ times when $L$ is increased from $100$~nm to $2\,\mu m$. However, we observe only modest variations of $F$ and $B$, inconsistent with local dissipation mechanism and suggesting the existence of a transport length scale comparable to the lengths of studied samples~\cite{manna2023}. We note that a possible contribution of non-thermal phonons does not affect this qualitative argument.\cite{Chen2020}

Our measurements show a significant temperature dependence of shot noise suppression. For instance, for a $500$~nm-long $\beta$-Ta wire $F = 0.14$ at 4.2 K, decreasing by a factor of $3$ to 0.05 at $T = 10$~K. This strong temperature dependence closely resembles the behaviors reported for a strange metal YbRh$_2$Si$_2${~\cite{chen2023shot}. For a $660$~nm-long nanowire, the value of $F$ interpolated from the published data for YbRh$_2$Si$_2$ is $0.1$ at $T=4$~K, decreasing by a factor of $5$ to $0.02$ at $10$~K. Furthermore, in both cases the e-ph coupling strength would have to dramatically increase in short wires to explain the length dependence of $F$.

Ref.{~\cite{chen2023shot}} attributed noise suppression in a strange metal to a continuous electron liquid which carries current but does not contribute to SN. Our results do not provide direct evidence for such a correlated electron liquid in $\beta$-Ta. Nevertheless, this possibility may account for its bad metal properties. A many-electron liquid cannot be characterized by the Bloch momentum of single-electron states, so its dynamics can result in energy transfer to the lattice even without lattice defects or phonons. This may explain the high resistivity that does not show the variation expected from  temperature-dependent e-ph scattering.}

Prior optical studies of disordered $\beta$-Ta provided additional evidence for many-electron effects. Increasing disorder resulted in a re-distribution of electron spectral weight from Drude contribution (free carriers) to the Lorentz contribution in the energy range of Mott-Hubbard transitions, suggesting an increasing role of correlations{~\cite{Kovaleva2015}}. Additionally, magnetic nano-islands embedded in $\beta$-Ta were shown to have a large effect on electron localization, indicating interplay between magnetism and localization attributed to quantum many-body phenomena{~\cite{Kovaleva2017,Kovaleva2020}}. A similar spectral weight redistribution was observed in temperature-dependent optical measurements of a strange metal Tb$_2$PdSi$_3$, which was attributed to electron trapping by magnetic clusters formed at low temperatures{~\cite{KovalevaTb}}. A similar correlation mechanism was proposed to explain localization effects due to magnetic nano-islands in disordered Ta{~\cite{Kovaleva2017}}.

%The usual single-electron SN picture or the ``hot" electron approximation describing electron gas in terms of the local effective temperature are not applicable to such a state. Extrapolating the dependence of $F$ on $T$ below the base temperature $T=4.2$~K of our measurement, one may expect anomalously large Fano factor at $T=0$, which would be consistent with transport of charge $q>e$. 

A rapid decrease of $F$ with increasing $T$ may result from the increased many-electron phase space provided by thermal energy, resulting in increasingly continuous electron liquid flow under bias. This may also explain the decrease of $\rho$ with $T$. The noise produced by such a system is influenced by $T_{ph}$, which cannot be described in terms of the single-electron temperature distribution given by Eq.~(\ref{eq:diffusion}).
%consistent with large discrepancies between data and calculations in Fig.~\ref{fig:4}(a). 

Our findings open several avenues for the exploration of correlated electron states in $\beta$-Ta and other anomalous metals, while also providing an efficient method of shot noise measurement for metallic nanojunctions. Studies at lower $T$ inaccessible in our measurements will likely reveal noise regimes that unambiguously confirm the non-Fermi liquid nature of electron state. The properties of this state can be further elucidated by the studies of SN in tunnel junctions. THz spectroscopy may reveal the characteristic energy scale involved in electron correlations. If the reported properties are confirmed for superconducting bad metals such as Mo$_x$Ge$_{1-x}$ and $\beta$-W, their SC may be associated with transition from viscous to superfluid state of the correlated electron liquid. Studies of nanojunctions formed by other bad metals may provide further insight into non-Fermi liquid electronic properties close to the localization limit. Their studies may shed light on the mechanisms of unconventional SC in other anomalous metals including high-temperature superconductors. Finally, the large SHE exhibited by $\beta$-Ta suggests the possibility to develop efficient spintronic devices based on correlated materials.

The authors have no conflicts to disclose.

Author Contributions: S.U. and Y.Z. conceived the experiments, M.S., H.C., and Z.P. fabricated the samples and performed measurements, J.B performed x-ray characterization. All authors contributed to data analysis and manuscript preparation.

We acknowledge support by the NSF award ECCS-2005786, the Tarbutton fellowship from Emory College of Arts and Sciences, and the SEED award from the Research Corporation for Science Advancement.

\section*{Data Availability Statement}
The data that support the findings of this study are available from the corresponding author upon reasonable request.

%\nocite{*}
\bibliography{Ta_Mat}

%aipnum4-2.bst 2019-01-14 (MD) hand-edited version of apsrev4-1.bst
\providecommand{\noopsort}[1]{}\providecommand{\singleletter}[1]{#1}%
\begin{thebibliography}{36}%
\makeatletter
\providecommand \@ifxundefined [1]{%
 \@ifx{#1\undefined}
}%
\providecommand \@ifnum [1]{%
 \ifnum #1\expandafter \@firstoftwo
 \else \expandafter \@secondoftwo
 \fi
}%
\providecommand \@ifx [1]{%
 \ifx #1\expandafter \@firstoftwo
 \else \expandafter \@secondoftwo
 \fi
}%
\providecommand \natexlab [1]{#1}%
\providecommand \enquote  [1]{``#1''}%
\providecommand \bibnamefont  [1]{#1}%
\providecommand \bibfnamefont [1]{#1}%
\providecommand \citenamefont [1]{#1}%
\providecommand \href@noop [0]{\@secondoftwo}%
\providecommand \href [0]{\begingroup \@sanitize@url \@href}%
\providecommand \@href[1]{\@@startlink{#1}\@@href}%
\providecommand \@@href[1]{\endgroup#1\@@endlink}%
\providecommand \@sanitize@url [0]{\catcode `\\12\catcode `\$12\catcode
  `\&12\catcode `\#12\catcode `\^12\catcode `\_12\catcode `\%12\relax}%
\providecommand \@@startlink[1]{}%
\providecommand \@@endlink[0]{}%
\providecommand \url  [0]{\begingroup\@sanitize@url \@url }%
\providecommand \@url [1]{\endgroup\@href {#1}{\urlprefix }}%
\providecommand \urlprefix  [0]{URL }%
\providecommand \Eprint [0]{\href }%
\providecommand \doibase [0]{https://doi.org/}%
\providecommand \selectlanguage [0]{\@gobble}%
\providecommand \bibinfo  [0]{\@secondoftwo}%
\providecommand \bibfield  [0]{\@secondoftwo}%
\providecommand \translation [1]{[#1]}%
\providecommand \BibitemOpen [0]{}%
\providecommand \bibitemStop [0]{}%
\providecommand \bibitemNoStop [0]{.\EOS\space}%
\providecommand \EOS [0]{\spacefactor3000\relax}%
\providecommand \BibitemShut  [1]{\csname bibitem#1\endcsname}%
\let\auto@bib@innerbib\@empty
%</preamble>
\bibitem [{\citenamefont {Kapitulnik}, \citenamefont {Kivelson},\ and\
  \citenamefont {Spivak}(2019)}]{RevModPhys.91.011002}%
  \BibitemOpen
  \bibfield  {author} {\bibinfo {author} {\bibfnamefont {A.}~\bibnamefont
  {Kapitulnik}}, \bibinfo {author} {\bibfnamefont {S.~A.}\ \bibnamefont
  {Kivelson}},\ and\ \bibinfo {author} {\bibfnamefont {B.}~\bibnamefont
  {Spivak}},\ }\bibfield  {title} {\enquote {\bibinfo {title} {Colloquium:
  Anomalous metals: Failed superconductors},}\ }\href
  {https://doi.org/10.1103/RevModPhys.91.011002} {\bibfield  {journal}
  {\bibinfo  {journal} {Rev. Mod. Phys.}\ }\textbf {\bibinfo {volume} {91}},\
  \bibinfo {pages} {011002} (\bibinfo {year} {2019})}\BibitemShut {NoStop}%
\bibitem [{\citenamefont {Chowdhury}\ \emph {et~al.}(2022)\citenamefont
  {Chowdhury}, \citenamefont {Georges}, \citenamefont {Parcollet},\ and\
  \citenamefont {Sachdev}}]{Chowdhury2022}%
  \BibitemOpen
  \bibfield  {author} {\bibinfo {author} {\bibfnamefont {D.}~\bibnamefont
  {Chowdhury}}, \bibinfo {author} {\bibfnamefont {A.}~\bibnamefont {Georges}},
  \bibinfo {author} {\bibfnamefont {O.}~\bibnamefont {Parcollet}},\ and\
  \bibinfo {author} {\bibfnamefont {S.}~\bibnamefont {Sachdev}},\ }\bibfield
  {title} {\enquote {\bibinfo {title} {{Sachdev-Ye-Kitaev models and beyond:
  Window into non-Fermi liquids}},}\ }\href@noop {} {\bibfield  {journal}
  {\bibinfo  {journal} {Reviews of Modern Physics}\ }\textbf {\bibinfo {volume}
  {94}},\ \bibinfo {pages} {035004} (\bibinfo {year} {2022})}\BibitemShut
  {NoStop}%
\bibitem [{\citenamefont {Else}\ and\ \citenamefont
  {Senthil}(2021)}]{PhysRevLett.127.086601}%
  \BibitemOpen
  \bibfield  {author} {\bibinfo {author} {\bibfnamefont {D.~V.}\ \bibnamefont
  {Else}}\ and\ \bibinfo {author} {\bibfnamefont {T.}~\bibnamefont {Senthil}},\
  }\bibfield  {title} {\enquote {\bibinfo {title} {{Strange Metals as Ersatz
  Fermi Liquids}},}\ }\href {https://doi.org/10.1103/PhysRevLett.127.086601}
  {\bibfield  {journal} {\bibinfo  {journal} {Phys. Rev. Lett.}\ }\textbf
  {\bibinfo {volume} {127}},\ \bibinfo {pages} {086601} (\bibinfo {year}
  {2021})}\BibitemShut {NoStop}%
\bibitem [{\citenamefont {Greene}(2023)}]{Greene2023}%
  \BibitemOpen
  \bibfield  {author} {\bibinfo {author} {\bibfnamefont {R.~L.}\ \bibnamefont
  {Greene}},\ }\bibfield  {title} {\enquote {\bibinfo {title} {{The strange
  metal state of the high-Tc cuprates}},}\ }\href
  {https://doi.org/10.1016/j.physc.2023.1354319} {\bibfield  {journal}
  {\bibinfo  {journal} {Physica C: Superconductivity and its Applications}\
  }\textbf {\bibinfo {volume} {612}},\ \bibinfo {pages} {1354319} (\bibinfo
  {year} {2023})}\BibitemShut {NoStop}%
\bibitem [{\citenamefont {Metlitski}\ \emph {et~al.}(2015)\citenamefont
  {Metlitski}, \citenamefont {Mross}, \citenamefont {Sachdev},\ and\
  \citenamefont {Senthil}}]{PhysRevB.91.115111}%
  \BibitemOpen
  \bibfield  {author} {\bibinfo {author} {\bibfnamefont {M.~A.}\ \bibnamefont
  {Metlitski}}, \bibinfo {author} {\bibfnamefont {D.~F.}\ \bibnamefont
  {Mross}}, \bibinfo {author} {\bibfnamefont {S.}~\bibnamefont {Sachdev}},\
  and\ \bibinfo {author} {\bibfnamefont {T.}~\bibnamefont {Senthil}},\
  }\bibfield  {title} {\enquote {\bibinfo {title} {{Cooper pairing in non-Fermi
  liquids}},}\ }\href {https://doi.org/10.1103/PhysRevB.91.115111} {\bibfield
  {journal} {\bibinfo  {journal} {Phys. Rev. B}\ }\textbf {\bibinfo {volume}
  {91}},\ \bibinfo {pages} {115111} (\bibinfo {year} {2015})}\BibitemShut
  {NoStop}%
\bibitem [{\citenamefont {Hegg}, \citenamefont {Hou},\ and\ \citenamefont
  {Ku}(2021)}]{Hegg2021}%
  \BibitemOpen
  \bibfield  {author} {\bibinfo {author} {\bibfnamefont {A.}~\bibnamefont
  {Hegg}}, \bibinfo {author} {\bibfnamefont {J.}~\bibnamefont {Hou}},\ and\
  \bibinfo {author} {\bibfnamefont {W.}~\bibnamefont {Ku}},\ }\bibfield
  {title} {\enquote {\bibinfo {title} {{Geometric frustration produces
  long-sought Bose metal phase of quantum matter}},}\ }\href@noop {} {\bibfield
   {journal} {\bibinfo  {journal} {Proceedings of the National Academy of
  Sciences}\ }\textbf {\bibinfo {volume} {118}},\ \bibinfo {pages}
  {e2100545118} (\bibinfo {year} {2021})}\BibitemShut {NoStop}%
\bibitem [{\citenamefont {Shklovskii}\ and\ \citenamefont
  {Efros}(1984)}]{shklovskii1984variable}%
  \BibitemOpen
  \bibfield  {author} {\bibinfo {author} {\bibfnamefont {B.~I.}\ \bibnamefont
  {Shklovskii}}\ and\ \bibinfo {author} {\bibfnamefont {A.~L.}\ \bibnamefont
  {Efros}},\ }\enquote {\bibinfo {title} {Variable-range hopping conduction},}\
  in\ \href {https://doi.org/10.1007/978-3-662-02403-4_9} {\emph {\bibinfo
  {booktitle} {Electronic Properties of Doped Semiconductors}}}\ (\bibinfo
  {publisher} {Springer Berlin Heidelberg},\ \bibinfo {address} {Berlin,
  Heidelberg},\ \bibinfo {year} {1984})\ pp.\ \bibinfo {pages}
  {202--227}\BibitemShut {NoStop}%
\bibitem [{\citenamefont {Zhang}, \citenamefont {Palevski},\ and\ \citenamefont
  {Kapitulnik}(2022)}]{Zhang2022}%
  \BibitemOpen
  \bibfield  {author} {\bibinfo {author} {\bibfnamefont {X.}~\bibnamefont
  {Zhang}}, \bibinfo {author} {\bibfnamefont {A.}~\bibnamefont {Palevski}},\
  and\ \bibinfo {author} {\bibfnamefont {A.}~\bibnamefont {Kapitulnik}},\
  }\bibfield  {title} {\enquote {\bibinfo {title} {Anomalous metals: From
  “failed superconductor” to “failed insulator”},}\ }\href@noop {}
  {\bibfield  {journal} {\bibinfo  {journal} {Proceedings of the National
  Academy of Sciences}\ }\textbf {\bibinfo {volume} {119}},\ \bibinfo {pages}
  {e2202496119} (\bibinfo {year} {2022})}\BibitemShut {NoStop}%
\bibitem [{\citenamefont {Mandal}\ \emph {et~al.}(2020)\citenamefont {Mandal},
  \citenamefont {Dutta}, \citenamefont {Basistha}, \citenamefont {Roy},
  \citenamefont {Jesudasan}, \citenamefont {Bagwe}, \citenamefont {Benfatto},
  \citenamefont {Thamizhavel},\ and\ \citenamefont
  {Raychaudhuri}}]{PhysRevB.102.060501}%
  \BibitemOpen
  \bibfield  {author} {\bibinfo {author} {\bibfnamefont {S.}~\bibnamefont
  {Mandal}}, \bibinfo {author} {\bibfnamefont {S.}~\bibnamefont {Dutta}},
  \bibinfo {author} {\bibfnamefont {S.}~\bibnamefont {Basistha}}, \bibinfo
  {author} {\bibfnamefont {I.}~\bibnamefont {Roy}}, \bibinfo {author}
  {\bibfnamefont {J.}~\bibnamefont {Jesudasan}}, \bibinfo {author}
  {\bibfnamefont {V.}~\bibnamefont {Bagwe}}, \bibinfo {author} {\bibfnamefont
  {L.}~\bibnamefont {Benfatto}}, \bibinfo {author} {\bibfnamefont
  {A.}~\bibnamefont {Thamizhavel}},\ and\ \bibinfo {author} {\bibfnamefont
  {P.}~\bibnamefont {Raychaudhuri}},\ }\bibfield  {title} {\enquote {\bibinfo
  {title} {{Destruction of superconductivity through phase fluctuations in
  ultrathin $a$-MoGe films}},}\ }\href
  {https://doi.org/10.1103/PhysRevB.102.060501} {\bibfield  {journal} {\bibinfo
   {journal} {Phys. Rev. B}\ }\textbf {\bibinfo {volume} {102}},\ \bibinfo
  {pages} {060501} (\bibinfo {year} {2020})}\BibitemShut {NoStop}%
\bibitem [{\citenamefont {Bastiaans}\ \emph {et~al.}(2021)\citenamefont
  {Bastiaans}, \citenamefont {Chatzopoulos}, \citenamefont {Ge}, \citenamefont
  {Cho}, \citenamefont {Tromp}, \citenamefont {van Ruitenbeek}, \citenamefont
  {Fischer}, \citenamefont {de~Visser}, \citenamefont {Thoen}, \citenamefont
  {Driessen}, \citenamefont {Klapwijk},\ and\ \citenamefont
  {Allan}}]{Bastiaans2021}%
  \BibitemOpen
  \bibfield  {author} {\bibinfo {author} {\bibfnamefont {K.~M.}\ \bibnamefont
  {Bastiaans}}, \bibinfo {author} {\bibfnamefont {D.}~\bibnamefont
  {Chatzopoulos}}, \bibinfo {author} {\bibfnamefont {J.-F.}\ \bibnamefont
  {Ge}}, \bibinfo {author} {\bibfnamefont {D.}~\bibnamefont {Cho}}, \bibinfo
  {author} {\bibfnamefont {W.~O.}\ \bibnamefont {Tromp}}, \bibinfo {author}
  {\bibfnamefont {J.~M.}\ \bibnamefont {van Ruitenbeek}}, \bibinfo {author}
  {\bibfnamefont {M.~H.}\ \bibnamefont {Fischer}}, \bibinfo {author}
  {\bibfnamefont {P.~J.}\ \bibnamefont {de~Visser}}, \bibinfo {author}
  {\bibfnamefont {D.~J.}\ \bibnamefont {Thoen}}, \bibinfo {author}
  {\bibfnamefont {E.~F.~C.}\ \bibnamefont {Driessen}}, \bibinfo {author}
  {\bibfnamefont {T.~M.}\ \bibnamefont {Klapwijk}},\ and\ \bibinfo {author}
  {\bibfnamefont {M.~P.}\ \bibnamefont {Allan}},\ }\bibfield  {title} {\enquote
  {\bibinfo {title} {{Direct evidence for Cooper pairing without a spectral gap
  in a disordered superconductor above T$_c$}},}\ }\href
  {https://doi.org/10.1126/science.abe3987} {\bibfield  {journal} {\bibinfo
  {journal} {Science}\ }\textbf {\bibinfo {volume} {374}},\ \bibinfo {pages}
  {608--611} (\bibinfo {year} {2021})}\BibitemShut {NoStop}%
\bibitem [{\citenamefont {Chauhan}, \citenamefont {Budhani},\ and\
  \citenamefont {Armitage}(2022)}]{PhysRevB.105.L060503}%
  \BibitemOpen
  \bibfield  {author} {\bibinfo {author} {\bibfnamefont {P.}~\bibnamefont
  {Chauhan}}, \bibinfo {author} {\bibfnamefont {R.}~\bibnamefont {Budhani}},\
  and\ \bibinfo {author} {\bibfnamefont {N.~P.}\ \bibnamefont {Armitage}},\
  }\bibfield  {title} {\enquote {\bibinfo {title} {Anomalously small
  superconducting gap in a strong spin-orbit coupled superconductor:
  $\beta$-tungsten},}\ }\href {https://doi.org/10.1103/PhysRevB.105.L060503}
  {\bibfield  {journal} {\bibinfo  {journal} {Phys. Rev. B}\ }\textbf {\bibinfo
  {volume} {105}},\ \bibinfo {pages} {L060503} (\bibinfo {year}
  {2022})}\BibitemShut {NoStop}%
\bibitem [{\citenamefont {Liu}\ \emph {et~al.}(2012)\citenamefont {Liu},
  \citenamefont {Pai}, \citenamefont {Li}, \citenamefont {Tseng}, \citenamefont
  {Ralph},\ and\ \citenamefont {Buhrman}}]{Liu555}%
  \BibitemOpen
  \bibfield  {author} {\bibinfo {author} {\bibfnamefont {L.}~\bibnamefont
  {Liu}}, \bibinfo {author} {\bibfnamefont {C.-F.}\ \bibnamefont {Pai}},
  \bibinfo {author} {\bibfnamefont {Y.}~\bibnamefont {Li}}, \bibinfo {author}
  {\bibfnamefont {H.~W.}\ \bibnamefont {Tseng}}, \bibinfo {author}
  {\bibfnamefont {D.~C.}\ \bibnamefont {Ralph}},\ and\ \bibinfo {author}
  {\bibfnamefont {R.~A.}\ \bibnamefont {Buhrman}},\ }\bibfield  {title}
  {\enquote {\bibinfo {title} {{Spin-Torque Switching with the Giant Spin Hall
  Effect of Tantalum}},}\ }\href {https://doi.org/10.1126/science.1218197}
  {\bibfield  {journal} {\bibinfo  {journal} {Science}\ }\textbf {\bibinfo
  {volume} {336}},\ \bibinfo {pages} {555--558} (\bibinfo {year}
  {2012})}\BibitemShut {NoStop}%
\bibitem [{\citenamefont {Wang}\ \emph {et~al.}(2022)\citenamefont {Wang},
  \citenamefont {Setty}, \citenamefont {Sur}, \citenamefont {Chen},
  \citenamefont {Paschen}, \citenamefont {Natelson},\ and\ \citenamefont
  {Si}}]{SNCorrelations}%
  \BibitemOpen
  \bibfield  {author} {\bibinfo {author} {\bibfnamefont {Y.}~\bibnamefont
  {Wang}}, \bibinfo {author} {\bibfnamefont {C.}~\bibnamefont {Setty}},
  \bibinfo {author} {\bibfnamefont {S.}~\bibnamefont {Sur}}, \bibinfo {author}
  {\bibfnamefont {L.}~\bibnamefont {Chen}}, \bibinfo {author} {\bibfnamefont
  {S.}~\bibnamefont {Paschen}}, \bibinfo {author} {\bibfnamefont
  {D.}~\bibnamefont {Natelson}},\ and\ \bibinfo {author} {\bibfnamefont
  {Q.}~\bibnamefont {Si}},\ }\href {https://arxiv.org/abs/2211.11735} {\enquote
  {\bibinfo {title} {Shot noise as a characterization of strongly correlated
  metals},}\ } (\bibinfo {year} {2022}),\ \Eprint
  {https://arxiv.org/abs/2211.11735} {arXiv:2211.11735 [cond-mat.str-el]}
  \BibitemShut {NoStop}%
\bibitem [{\citenamefont {Nikolaenko}, \citenamefont {Sachdev},\ and\
  \citenamefont {Patel}(2023)}]{ShotNoiseStrangeMetals}%
  \BibitemOpen
  \bibfield  {author} {\bibinfo {author} {\bibfnamefont {A.}~\bibnamefont
  {Nikolaenko}}, \bibinfo {author} {\bibfnamefont {S.}~\bibnamefont
  {Sachdev}},\ and\ \bibinfo {author} {\bibfnamefont {A.~A.}\ \bibnamefont
  {Patel}},\ }\bibfield  {title} {\enquote {\bibinfo {title} {Theory of shot
  noise in strange metals},}\ }\href
  {https://doi.org/10.1103/PhysRevResearch.5.043143} {\bibfield  {journal}
  {\bibinfo  {journal} {Phys. Rev. Res.}\ }\textbf {\bibinfo {volume} {5}},\
  \bibinfo {pages} {043143} (\bibinfo {year} {2023})}\BibitemShut {NoStop}%
\bibitem [{\citenamefont {Chen}\ \emph {et~al.}(2023)\citenamefont {Chen},
  \citenamefont {Lowder}, \citenamefont {Bakali}, \citenamefont {Andrews},
  \citenamefont {Schrenk}, \citenamefont {Waas}, \citenamefont {Svagera},
  \citenamefont {Eguchi}, \citenamefont {Prochaska}, \citenamefont {Wang} \emph
  {et~al.}}]{chen2023shot}%
  \BibitemOpen
  \bibfield  {author} {\bibinfo {author} {\bibfnamefont {L.}~\bibnamefont
  {Chen}}, \bibinfo {author} {\bibfnamefont {D.~T.}\ \bibnamefont {Lowder}},
  \bibinfo {author} {\bibfnamefont {E.}~\bibnamefont {Bakali}}, \bibinfo
  {author} {\bibfnamefont {A.~M.}\ \bibnamefont {Andrews}}, \bibinfo {author}
  {\bibfnamefont {W.}~\bibnamefont {Schrenk}}, \bibinfo {author} {\bibfnamefont
  {M.}~\bibnamefont {Waas}}, \bibinfo {author} {\bibfnamefont {R.}~\bibnamefont
  {Svagera}}, \bibinfo {author} {\bibfnamefont {G.}~\bibnamefont {Eguchi}},
  \bibinfo {author} {\bibfnamefont {L.}~\bibnamefont {Prochaska}}, \bibinfo
  {author} {\bibfnamefont {Y.}~\bibnamefont {Wang}}, \emph {et~al.},\
  }\bibfield  {title} {\enquote {\bibinfo {title} {Shot noise in a strange
  metal},}\ }\href@noop {} {\bibfield  {journal} {\bibinfo  {journal}
  {Science}\ }\textbf {\bibinfo {volume} {382}},\ \bibinfo {pages} {907--911}
  (\bibinfo {year} {2023})}\BibitemShut {NoStop}%
\bibitem [{\citenamefont {Wu}\ and\ \citenamefont
  {Foster}(2024)}]{PhysRevB.110.L081102}%
  \BibitemOpen
  \bibfield  {author} {\bibinfo {author} {\bibfnamefont {T.~C.}\ \bibnamefont
  {Wu}}\ and\ \bibinfo {author} {\bibfnamefont {M.~S.}\ \bibnamefont
  {Foster}},\ }\bibfield  {title} {\enquote {\bibinfo {title} {{Suppression of
  shot noise in a dirty marginal Fermi liquid}},}\ }\href
  {https://doi.org/10.1103/PhysRevB.110.L081102} {\bibfield  {journal}
  {\bibinfo  {journal} {Phys. Rev. B}\ }\textbf {\bibinfo {volume} {110}},\
  \bibinfo {pages} {L081102} (\bibinfo {year} {2024})}\BibitemShut {NoStop}%
\bibitem [{\citenamefont {Zhang}\ \emph {et~al.}(2024)\citenamefont {Zhang},
  \citenamefont {Pandey}, \citenamefont {Ivanov}, \citenamefont {Liu},\ and\
  \citenamefont {Urazhdin}}]{SNYiou}%
  \BibitemOpen
  \bibfield  {author} {\bibinfo {author} {\bibfnamefont {Y.}~\bibnamefont
  {Zhang}}, \bibinfo {author} {\bibfnamefont {S.}~\bibnamefont {Pandey}},
  \bibinfo {author} {\bibfnamefont {S.}~\bibnamefont {Ivanov}}, \bibinfo
  {author} {\bibfnamefont {J.}~\bibnamefont {Liu}},\ and\ \bibinfo {author}
  {\bibfnamefont {S.}~\bibnamefont {Urazhdin}},\ }\bibfield  {title} {\enquote
  {\bibinfo {title} {{Shot noise in a metal close to Mott transition}},}\
  }\href {https://arxiv.org/abs/2405.18704} {\bibfield  {journal} {\bibinfo
  {journal} {arXiv:2405.18704}\ } (\bibinfo {year} {2024})}\BibitemShut
  {NoStop}%
\bibitem [{\citenamefont {Blanter}\ and\ \citenamefont
  {Büttiker}(2000)}]{ShotNoiseBible}%
  \BibitemOpen
  \bibfield  {author} {\bibinfo {author} {\bibfnamefont {Y.}~\bibnamefont
  {Blanter}}\ and\ \bibinfo {author} {\bibfnamefont {M.}~\bibnamefont
  {Büttiker}},\ }\bibfield  {title} {\enquote {\bibinfo {title} {Shot noise in
  mesoscopic conductors},}\ }\href
  {https://doi.org/https://doi.org/10.1016/S0370-1573(99)00123-4} {\bibfield
  {journal} {\bibinfo  {journal} {Physics Reports}\ }\textbf {\bibinfo {volume}
  {336}},\ \bibinfo {pages} {1--166} (\bibinfo {year} {2000})}\BibitemShut
  {NoStop}%
\bibitem [{\citenamefont {Beenakker}\ and\ \citenamefont
  {B{\"u}ttiker}(1992)}]{beenakker1992suppression}%
  \BibitemOpen
  \bibfield  {author} {\bibinfo {author} {\bibfnamefont {C.}~\bibnamefont
  {Beenakker}}\ and\ \bibinfo {author} {\bibfnamefont {M.}~\bibnamefont
  {B{\"u}ttiker}},\ }\bibfield  {title} {\enquote {\bibinfo {title}
  {Suppression of shot noise in metallic diffusive conductors},}\ }\href@noop
  {} {\bibfield  {journal} {\bibinfo  {journal} {Physical Review B}\ }\textbf
  {\bibinfo {volume} {46}},\ \bibinfo {pages} {1889} (\bibinfo {year}
  {1992})}\BibitemShut {NoStop}%
\bibitem [{\citenamefont {Nagaev}(1992)}]{SNDirtyMetal}%
  \BibitemOpen
  \bibfield  {author} {\bibinfo {author} {\bibfnamefont {K.}~\bibnamefont
  {Nagaev}},\ }\bibfield  {title} {\enquote {\bibinfo {title} {On the shot
  noise in dirty metal contacts},}\ }\href
  {https://doi.org/https://doi.org/10.1016/0375-9601(92)90814-3} {\bibfield
  {journal} {\bibinfo  {journal} {Physics Letters A}\ }\textbf {\bibinfo
  {volume} {169}},\ \bibinfo {pages} {103--107} (\bibinfo {year}
  {1992})}\BibitemShut {NoStop}%
\bibitem [{\citenamefont {Steinbach}, \citenamefont {Martinis},\ and\
  \citenamefont {Devoret}(1996)}]{HotElectrons}%
  \BibitemOpen
  \bibfield  {author} {\bibinfo {author} {\bibfnamefont {A.~H.}\ \bibnamefont
  {Steinbach}}, \bibinfo {author} {\bibfnamefont {J.~M.}\ \bibnamefont
  {Martinis}},\ and\ \bibinfo {author} {\bibfnamefont {M.~H.}\ \bibnamefont
  {Devoret}},\ }\bibfield  {title} {\enquote {\bibinfo {title} {Observation of
  hot-electron shot noise in a metallic resistor},}\ }\href
  {https://doi.org/10.1103/PhysRevLett.76.3806} {\bibfield  {journal} {\bibinfo
   {journal} {Phys. Rev. Lett.}\ }\textbf {\bibinfo {volume} {76}},\ \bibinfo
  {pages} {3806--3809} (\bibinfo {year} {1996})}\BibitemShut {NoStop}%
\bibitem [{\citenamefont {Nagaev}(1995)}]{SNelectronElectron}%
  \BibitemOpen
  \bibfield  {author} {\bibinfo {author} {\bibfnamefont {K.~E.}\ \bibnamefont
  {Nagaev}},\ }\bibfield  {title} {\enquote {\bibinfo {title} {Influence of
  electron-electron scattering on shot noise in diffusive contacts},}\ }\href
  {https://doi.org/10.1103/PhysRevB.52.4740} {\bibfield  {journal} {\bibinfo
  {journal} {Phys. Rev. B}\ }\textbf {\bibinfo {volume} {52}},\ \bibinfo
  {pages} {4740--4743} (\bibinfo {year} {1995})}\BibitemShut {NoStop}%
\bibitem [{not()}]{note_hot_e}%
  \BibitemOpen
  \href@noop {} {}\bibinfo {note} {In this regime $coth(V/V_{th})$ is an
  approximation for $tan^{-1}(\sqrt{3}V/\pi V_{th}$)}\BibitemShut {NoStop}%
\bibitem [{\citenamefont {Wellstood}, \citenamefont {Urbina},\ and\
  \citenamefont {Clarke}(1994)}]{PhysRevB.49.5942}%
  \BibitemOpen
  \bibfield  {author} {\bibinfo {author} {\bibfnamefont {F.~C.}\ \bibnamefont
  {Wellstood}}, \bibinfo {author} {\bibfnamefont {C.}~\bibnamefont {Urbina}},\
  and\ \bibinfo {author} {\bibfnamefont {J.}~\bibnamefont {Clarke}},\
  }\bibfield  {title} {\enquote {\bibinfo {title} {Hot-electron effects in
  metals},}\ }\href {https://doi.org/10.1103/PhysRevB.49.5942} {\bibfield
  {journal} {\bibinfo  {journal} {Phys. Rev. B}\ }\textbf {\bibinfo {volume}
  {49}},\ \bibinfo {pages} {5942--5955} (\bibinfo {year} {1994})}\BibitemShut
  {NoStop}%
\bibitem [{\citenamefont {Kinkhabwala}\ \emph {et~al.}(2006)\citenamefont
  {Kinkhabwala}, \citenamefont {Sverdlov}, \citenamefont {Korotkov},\ and\
  \citenamefont {Likharev}}]{kinkhabwala2006numerical}%
  \BibitemOpen
  \bibfield  {author} {\bibinfo {author} {\bibfnamefont {Y.~A.}\ \bibnamefont
  {Kinkhabwala}}, \bibinfo {author} {\bibfnamefont {V.~A.}\ \bibnamefont
  {Sverdlov}}, \bibinfo {author} {\bibfnamefont {A.~N.}\ \bibnamefont
  {Korotkov}},\ and\ \bibinfo {author} {\bibfnamefont {K.~K.}\ \bibnamefont
  {Likharev}},\ }\bibfield  {title} {\enquote {\bibinfo {title} {A numerical
  study of transport and shot noise in 2d hopping},}\ }\href@noop {} {\bibfield
   {journal} {\bibinfo  {journal} {Journal of Physics: Condensed Matter}\
  }\textbf {\bibinfo {volume} {18}},\ \bibinfo {pages} {1999} (\bibinfo {year}
  {2006})}\BibitemShut {NoStop}%
\bibitem [{\citenamefont {Ellis}, \citenamefont {Chmielus},\ and\ \citenamefont
  {Baker}(2018)}]{bTaProduction}%
  \BibitemOpen
  \bibfield  {author} {\bibinfo {author} {\bibfnamefont {E.~A.}\ \bibnamefont
  {Ellis}}, \bibinfo {author} {\bibfnamefont {M.}~\bibnamefont {Chmielus}},\
  and\ \bibinfo {author} {\bibfnamefont {S.~P.}\ \bibnamefont {Baker}},\
  }\bibfield  {title} {\enquote {\bibinfo {title} {{Effect of sputter pressure
  on Ta thin films: Beta phase formation, texture, and stresses}},}\ }\href
  {https://doi.org/https://doi.org/10.1016/j.actamat.2018.02.050} {\bibfield
  {journal} {\bibinfo  {journal} {Acta Materialia}\ }\textbf {\bibinfo {volume}
  {150}},\ \bibinfo {pages} {317--326} (\bibinfo {year} {2018})}\BibitemShut
  {NoStop}%
\bibitem [{\citenamefont {Jiang}\ \emph {et~al.}(2003)\citenamefont {Jiang},
  \citenamefont {Yohannan}, \citenamefont {Nnolim}, \citenamefont {Tyson},
  \citenamefont {Axe}, \citenamefont {Lee},\ and\ \citenamefont
  {Cote}}]{bTaStructure}%
  \BibitemOpen
  \bibfield  {author} {\bibinfo {author} {\bibfnamefont {A.}~\bibnamefont
  {Jiang}}, \bibinfo {author} {\bibfnamefont {A.}~\bibnamefont {Yohannan}},
  \bibinfo {author} {\bibfnamefont {N.~O.}\ \bibnamefont {Nnolim}}, \bibinfo
  {author} {\bibfnamefont {T.~A.}\ \bibnamefont {Tyson}}, \bibinfo {author}
  {\bibfnamefont {L.}~\bibnamefont {Axe}}, \bibinfo {author} {\bibfnamefont
  {S.~L.}\ \bibnamefont {Lee}},\ and\ \bibinfo {author} {\bibfnamefont
  {P.}~\bibnamefont {Cote}},\ }\bibfield  {title} {\enquote {\bibinfo {title}
  {Investigation of the structure of b-tantalum},}\ }\href
  {https://doi.org/https://doi.org/10.1016/S0040-6090(03)00702-8} {\bibfield
  {journal} {\bibinfo  {journal} {Thin Solid Films}\ }\textbf {\bibinfo
  {volume} {437}},\ \bibinfo {pages} {116--122} (\bibinfo {year}
  {2003})}\BibitemShut {NoStop}%
\bibitem [{\citenamefont {Magnuson}\ \emph {et~al.}(2019)\citenamefont
  {Magnuson}, \citenamefont {Greczynski}, \citenamefont {Eriksson},
  \citenamefont {Hultman},\ and\ \citenamefont {Hogberg}}]{Magnuson2019}%
  \BibitemOpen
  \bibfield  {author} {\bibinfo {author} {\bibfnamefont {M.}~\bibnamefont
  {Magnuson}}, \bibinfo {author} {\bibfnamefont {G.}~\bibnamefont
  {Greczynski}}, \bibinfo {author} {\bibfnamefont {F.}~\bibnamefont
  {Eriksson}}, \bibinfo {author} {\bibfnamefont {L.}~\bibnamefont {Hultman}},\
  and\ \bibinfo {author} {\bibfnamefont {H.}~\bibnamefont {Hogberg}},\
  }\bibfield  {title} {\enquote {\bibinfo {title} {{Electronic structure of
  $\beta$-Ta films from X-ray photoelectron spectroscopy and first-principles
  calculations}},}\ }\href {https://doi.org/10.1016/j.apsusc.2018.11.096}
  {\bibfield  {journal} {\bibinfo  {journal} {Applied Surface Science}\
  }\textbf {\bibinfo {volume} {470}},\ \bibinfo {pages} {607--612} (\bibinfo
  {year} {2019})}\BibitemShut {NoStop}%
\bibitem [{\citenamefont {Henny}\ \emph {et~al.}(1997)\citenamefont {Henny},
  \citenamefont {Birk}, \citenamefont {Huber}, \citenamefont {Strunk},
  \citenamefont {Bachtold}, \citenamefont {Kr\"{u}ger},\ and\ \citenamefont
  {Sch\"{o}nenberger}}]{Henny1997}%
  \BibitemOpen
  \bibfield  {author} {\bibinfo {author} {\bibfnamefont {M.}~\bibnamefont
  {Henny}}, \bibinfo {author} {\bibfnamefont {H.}~\bibnamefont {Birk}},
  \bibinfo {author} {\bibfnamefont {R.}~\bibnamefont {Huber}}, \bibinfo
  {author} {\bibfnamefont {C.}~\bibnamefont {Strunk}}, \bibinfo {author}
  {\bibfnamefont {A.}~\bibnamefont {Bachtold}}, \bibinfo {author}
  {\bibfnamefont {M.}~\bibnamefont {Kr\"{u}ger}},\ and\ \bibinfo {author}
  {\bibfnamefont {C.}~\bibnamefont {Sch\"{o}nenberger}},\ }\bibfield  {title}
  {\enquote {\bibinfo {title} {Electron heating effects in diffusive metal
  wires},}\ }\href {https://doi.org/10.1063/1.119641} {\bibfield  {journal}
  {\bibinfo  {journal} {Applied Physics Letters}\ }\textbf {\bibinfo {volume}
  {71}},\ \bibinfo {pages} {773–775} (\bibinfo {year} {1997})}\BibitemShut
  {NoStop}%
\bibitem [{\citenamefont {Naveh}, \citenamefont {Averin},\ and\ \citenamefont
  {Likharev}(1998)}]{SNDiffusive}%
  \BibitemOpen
  \bibfield  {author} {\bibinfo {author} {\bibfnamefont {Y.}~\bibnamefont
  {Naveh}}, \bibinfo {author} {\bibfnamefont {D.~V.}\ \bibnamefont {Averin}},\
  and\ \bibinfo {author} {\bibfnamefont {K.~K.}\ \bibnamefont {Likharev}},\
  }\bibfield  {title} {\enquote {\bibinfo {title} {Shot noise in diffusive
  conductors: A quantitative analysis of electron-phonon interaction
  effects},}\ }\href {https://doi.org/10.1103/physrevb.58.15371} {\bibfield
  {journal} {\bibinfo  {journal} {Physical Review B}\ }\textbf {\bibinfo
  {volume} {58}},\ \bibinfo {pages} {15371–15374} (\bibinfo {year}
  {1998})}\BibitemShut {NoStop}%
\bibitem [{\citenamefont {Manna}\ and\ \citenamefont {Das}(2023)}]{manna2023}%
  \BibitemOpen
  \bibfield  {author} {\bibinfo {author} {\bibfnamefont {S.}~\bibnamefont
  {Manna}}\ and\ \bibinfo {author} {\bibfnamefont {A.}~\bibnamefont {Das}},\
  }\bibfield  {title} {\enquote {\bibinfo {title} {Experimentally motivated
  order of length scales affect shot noise},}\ }\href
  {https://arxiv.org/abs/2307.08264} {\bibfield  {journal} {\bibinfo  {journal}
  {arXiv:2307.08264}\ } (\bibinfo {year} {2023})}\BibitemShut {NoStop}%
\bibitem [{\citenamefont {Chen}\ \emph {et~al.}(2020)\citenamefont {Chen},
  \citenamefont {Freeman}, \citenamefont {Zholud},\ and\ \citenamefont
  {Urazhdin}}]{Chen2020}%
  \BibitemOpen
  \bibfield  {author} {\bibinfo {author} {\bibfnamefont {G.}~\bibnamefont
  {Chen}}, \bibinfo {author} {\bibfnamefont {R.}~\bibnamefont {Freeman}},
  \bibinfo {author} {\bibfnamefont {A.}~\bibnamefont {Zholud}},\ and\ \bibinfo
  {author} {\bibfnamefont {S.}~\bibnamefont {Urazhdin}},\ }\bibfield  {title}
  {\enquote {\bibinfo {title} {Observation of anomalous non-ohmic transport in
  current-driven nanostructures},}\ }\href
  {https://doi.org/10.1103/physrevx.10.011064} {\bibfield  {journal} {\bibinfo
  {journal} {Physical Review X}\ }\textbf {\bibinfo {volume} {10}} (\bibinfo
  {year} {2020}),\ 10.1103/physrevx.10.011064}\BibitemShut {NoStop}%
\bibitem [{\citenamefont {Kovaleva}\ \emph {et~al.}(2015)\citenamefont
  {Kovaleva}, \citenamefont {Chvostova}, \citenamefont {Bagdinov},
  \citenamefont {Petrova}, \citenamefont {Demikhov}, \citenamefont {Pudonin},\
  and\ \citenamefont {Dejneka}}]{Kovaleva2015}%
  \BibitemOpen
  \bibfield  {author} {\bibinfo {author} {\bibfnamefont {N.~N.}\ \bibnamefont
  {Kovaleva}}, \bibinfo {author} {\bibfnamefont {D.}~\bibnamefont {Chvostova}},
  \bibinfo {author} {\bibfnamefont {A.~V.}\ \bibnamefont {Bagdinov}}, \bibinfo
  {author} {\bibfnamefont {M.~G.}\ \bibnamefont {Petrova}}, \bibinfo {author}
  {\bibfnamefont {E.~I.}\ \bibnamefont {Demikhov}}, \bibinfo {author}
  {\bibfnamefont {F.~A.}\ \bibnamefont {Pudonin}},\ and\ \bibinfo {author}
  {\bibfnamefont {A.}~\bibnamefont {Dejneka}},\ }\bibfield  {title} {\enquote
  {\bibinfo {title} {Interplay of electron correlations and localization in
  disordered $\beta$-tantalum films: Evidence from dc transport and
  spectroscopic ellipsometry study},}\ }\href
  {https://doi.org/10.1063/1.4907862} {\bibfield  {journal} {\bibinfo
  {journal} {Applied Physics Letters}\ }\textbf {\bibinfo {volume} {106}},\
  \bibinfo {pages} {051907} (\bibinfo {year} {2015})},\ \Eprint
  {https://arxiv.org/abs/https://pubs.aip.org/aip/apl/article-pdf/doi/10.1063/1.4907862/14464226/051907\_1\_online.pdf}
  {https://pubs.aip.org/aip/apl/article-pdf/doi/10.1063/1.4907862/14464226/051907\_1\_online.pdf}
  \BibitemShut {NoStop}%
\bibitem [{\citenamefont {Kovaleva}, \citenamefont {Chvostova},\ and\
  \citenamefont {Dejneka}(2017)}]{Kovaleva2017}%
  \BibitemOpen
  \bibfield  {author} {\bibinfo {author} {\bibfnamefont {N.}~\bibnamefont
  {Kovaleva}}, \bibinfo {author} {\bibfnamefont {D.}~\bibnamefont
  {Chvostova}},\ and\ \bibinfo {author} {\bibfnamefont {A.}~\bibnamefont
  {Dejneka}},\ }\bibfield  {title} {\enquote {\bibinfo {title} {Localization
  phenomena in disordered tantalum films},}\ }\href
  {https://doi.org/10.3390/met7070257} {\bibfield  {journal} {\bibinfo
  {journal} {Metals}\ }\textbf {\bibinfo {volume} {7}},\ \bibinfo {pages} {257}
  (\bibinfo {year} {2017})}\BibitemShut {NoStop}%
\bibitem [{\citenamefont {Kovaleva}\ \emph {et~al.}(2020)\citenamefont
  {Kovaleva}, \citenamefont {Kusmartsev}, \citenamefont {Mekhiya},
  \citenamefont {Trunkin}, \citenamefont {Chvostova}, \citenamefont {Davydov},
  \citenamefont {Oveshnikov}, \citenamefont {Pacherova}, \citenamefont
  {Sherstnev}, \citenamefont {Kusmartseva}, \citenamefont {Kugel},
  \citenamefont {Dejneka}, \citenamefont {Pudonin}, \citenamefont {Luo},\ and\
  \citenamefont {Aronzon}}]{Kovaleva2020}%
  \BibitemOpen
  \bibfield  {author} {\bibinfo {author} {\bibfnamefont {N.~N.}\ \bibnamefont
  {Kovaleva}}, \bibinfo {author} {\bibfnamefont {F.~V.}\ \bibnamefont
  {Kusmartsev}}, \bibinfo {author} {\bibfnamefont {A.~B.}\ \bibnamefont
  {Mekhiya}}, \bibinfo {author} {\bibfnamefont {I.~N.}\ \bibnamefont
  {Trunkin}}, \bibinfo {author} {\bibfnamefont {D.}~\bibnamefont {Chvostova}},
  \bibinfo {author} {\bibfnamefont {A.~B.}\ \bibnamefont {Davydov}}, \bibinfo
  {author} {\bibfnamefont {L.~N.}\ \bibnamefont {Oveshnikov}}, \bibinfo
  {author} {\bibfnamefont {O.}~\bibnamefont {Pacherova}}, \bibinfo {author}
  {\bibfnamefont {I.~A.}\ \bibnamefont {Sherstnev}}, \bibinfo {author}
  {\bibfnamefont {A.}~\bibnamefont {Kusmartseva}}, \bibinfo {author}
  {\bibfnamefont {K.~I.}\ \bibnamefont {Kugel}}, \bibinfo {author}
  {\bibfnamefont {A.}~\bibnamefont {Dejneka}}, \bibinfo {author} {\bibfnamefont
  {F.~A.}\ \bibnamefont {Pudonin}}, \bibinfo {author} {\bibfnamefont
  {Y.}~\bibnamefont {Luo}},\ and\ \bibinfo {author} {\bibfnamefont {B.~A.}\
  \bibnamefont {Aronzon}},\ }\bibfield  {title} {\enquote {\bibinfo {title}
  {Control of mooij correlations at the nanoscale in the disordered metallic
  ta–nanoisland feni multilayers},}\ }\href
  {https://doi.org/10.1038/s41598-020-78185-6} {\bibfield  {journal} {\bibinfo
  {journal} {Scientific Reports}\ }\textbf {\bibinfo {volume} {10}} (\bibinfo
  {year} {2020}),\ 10.1038/s41598-020-78185-6}\BibitemShut {NoStop}%
\bibitem [{\citenamefont {Kovaleva}\ \emph {et~al.}(2012)\citenamefont
  {Kovaleva}, \citenamefont {Kugel}, \citenamefont {Bazhenov}, \citenamefont
  {Fursova}, \citenamefont {L\"{o}ser}, \citenamefont {Xu}, \citenamefont
  {Behr},\ and\ \citenamefont {Kusmartsev}}]{KovalevaTb}%
  \BibitemOpen
  \bibfield  {author} {\bibinfo {author} {\bibfnamefont {N.~N.}\ \bibnamefont
  {Kovaleva}}, \bibinfo {author} {\bibfnamefont {K.~I.}\ \bibnamefont {Kugel}},
  \bibinfo {author} {\bibfnamefont {A.~V.}\ \bibnamefont {Bazhenov}}, \bibinfo
  {author} {\bibfnamefont {T.~N.}\ \bibnamefont {Fursova}}, \bibinfo {author}
  {\bibfnamefont {W.}~\bibnamefont {L\"{o}ser}}, \bibinfo {author}
  {\bibfnamefont {Y.}~\bibnamefont {Xu}}, \bibinfo {author} {\bibfnamefont
  {G.}~\bibnamefont {Behr}},\ and\ \bibinfo {author} {\bibfnamefont {F.~V.}\
  \bibnamefont {Kusmartsev}},\ }\bibfield  {title} {\enquote {\bibinfo {title}
  {Formation of metallic magnetic clusters in a kondo-lattice metal: Evidence
  from an optical study},}\ }\href {https://doi.org/10.1038/srep00890}
  {\bibfield  {journal} {\bibinfo  {journal} {Scientific Reports}\ }\textbf
  {\bibinfo {volume} {2}} (\bibinfo {year} {2012}),\
  10.1038/srep00890}\BibitemShut {NoStop}%
\end{thebibliography}%

\end{document}